\documentclass[12pt]{article}
\usepackage{graphicx}
\usepackage{amssymb}
\title{String model of the Hydrogen Atom }
\author{Omar Y\'epez\\
Department of Chemistry, Memorial University of Newfoundland, \\
St. JohnÕs, NF, A1B 3X7
Canada.}
\begin{document}
\maketitle
\begin{abstract}
A non-moving electron hydrogen model is proposed, resolving a long standing contradiction (94 years) in the hydrogen atom. This, however, forces to not use the "in an orbit point particle kinetic energy" as the phenomenon responsible for the atom stability. The repulsion between the masses of the electron and proton is what is responsible of such stability. The mass of the electron is a field fully described by the uncertainty principle through the confinement of the particle, which is also consistent with the general theory of relativity that states: "mass-energy tells the space how to bend". Ergo, mass exerts a tension on its surrounding space and the lighter the mass the larger the space it will occupy. Based on this concept it is proposed that the orbital is the electron. The electron's orbitals are just the electron's different ways of intersecting the space; with different magnetic momenta. The coupling of this momenta with the magnetic moment of the proton finally explains the hyperfine structure of the hydrogen spectrum with an overwhelming simplicity.\\

\end{abstract}
\section{Introduction}
The hydrogen atom, as it is currently understood, involves a non-radiating, moving charged point particle (the electron) in an orbit around the proton, where the orbit's  shape is dictated by the wave equation \cite{eisberg277}. Regardless of the orbit' size, the electron charge is spread out within the orbit. The electron mass is somehow contained in that point and the electron radiates when it jumps from one orbit to another but not while it is moving in it, which is contradictory. This is required to avoid energy losses and an eventual collapse into the proton. However, the product of such collapse, a neutron, requires more energy than the sum of the mass energies of the proton and the electron, which means that they can be motionless together and a neutron will not be produced \footnote{The mass difference between the neutron and the sum of the proton and electron is 0.78 MeV. This energy has to be paid to produce a neutron}. The energy required for the electron to move in this orbit has been completely ignored, as well as, how the electron while in this chaotic movement, copes with all of the shape of the orbit  (in dumb bell like orbits) and still produces a highly directional magnetic moment.\\
The critics to the orbit model begun with the non-radiating moving point charge which frontally violates the fact that a moving charge emits electromagnetic waves and continued with a  thermodynamic contradiction: the electron movement will necessarily produce an induced alternate current in a fine wire approached to the atom \cite{chemist electron}, i.e. a perpetual motion machine of the first kind. Another well documented contradiction is that a point particle with infinite energy will have an intrinsic charge instability.\\
Moreover, historically, successful theories for the hydrogen atom survived until the precision in the measurements of the hydrogen spectrum increased and, actually, no comprehensible hydrogen theory exists. The actual wave mechanical description of the hydrogen atom comes short to explain the fine and hyperfine structure of the hydrogen spectrum because its formalism does not include the existence of an electron magnetic moment. Dirac came up with a solution as the magnetic moment of the electron appears naturally in his treatment, being able to reproduce Sommerfeld equation (based on a model with elliptic orbits) and explaining the fine structure of the spectrum  successfully by invoking the spin-orbit interaction (not the elliptic orbits!), which is the coupling of the magnetic moment of the electron with the angular momentum of its own orbit. The Dirac model at the time failed when the hyperfine structure of the hydrogen spectrum was discovered and Lamb and Retherford demonstrated its inadequacy \cite{Hspectrum1}. 
Finally, the difficulties encountered by the Dirac theory were circumvented by quantum electrodynamics making small corrections in the energy levels predicted by Dirac's theory. This solution, however, is not a  complete and logically satisfactory solution to the problem of the hydrogen atom: a difficulty of principle remains now, as formerly \cite{Hspectrum1}. \\
By looking to the intersection of a torus by a plane, a completely new particle model that describes elementary particles as the intersection of a 4 dimensional torus into a 3 dimensional space has been described \cite{yepez, yepez2}. Thanks to the use of this higher dimensional view, unexplainable phenomena such as self -interference and the EPR paradox has been explained in a simpler way. The main characteristic of this model is that it involves the interaction between the particle and its surrounding space. This has made it easier to detect that in order to have mass, an electron uses all the dimensions of the space it is intersecting. The intersection events produce electric field vectors that tense the particle's surrounding space, precisely because such dimensions usage and finally producing the particle mass \cite{yepez}. The use of a higher dimensional object intersecting a lower dimensional space has been very productive in explaining why the position and momentum of the particle does not commute and in showing an intimate connection between the magnetic and the anapole moment of a particle: both properties respectively depend on how the space is intersected by the particle and that is why each pair of properties respectively do not commute. So far, this model had explained the charge, magnetic moment, and shapes experimentally found for nuclei \cite{forest}, as well as self-interference, number of rounds before reaching identity for the electron and photon, Stern-Gerlach experiments, EPR paradox, the appearance of the anapole moment and parity violation (chirality). Furthermore, this model also provides a possible concept (at least consistent with the general theory of relativity, GTR) for the mass of the particle \cite{yepez, yepez2}.\\
The "GTR consistent concept of mass" encounters its direct connection with Quantum Mechanics in the uncertainty principle which explicitly relates the energy of the particle (its mass energy) with its confinement, i.e. the lighter the object, the larger the space it will occupy. This opens the possibility for the electron to be just the orbital, avoiding the radiation problem because the orbit=electron is not moving.\\
The kinetic energy can not longer be offered to explain the atom stability. The uncertainty principle has been invoked to have the kinetic energy of the point electron in the orbit \cite{eisberg297}, but the same result can be obtained by the energy-time uncertainty relation where the energy is not necessarily kinetic. If the electron=orbit is not moving, such energy is proposed to be the potential energy due to the presence of the particle, i.e. its mass energy, which is repulsive between the two masses and works against the coulomb attraction. Therefore, the big energy that has to be paid in order to fuse the masses of the electron and the proton is what prevents the collapse, because the Coulomb attraction is not enough to produce it.\\
The energy term that comes from the uncertainty principle and which is inversely proportional to the square of the atom radius is due to the mass energy of the electron and decreases as the area it occupies increases \cite{eisberg297}, which is consistent with the general theory of relativity that states that mass tells the space how to bend \cite{hyperspace}, i.e.  mass exerts a tension on the space, therefore the smaller the particle mass, the larger the space it will occupy; thus, the electron can be the orbit.\\
The advantage of this hypothesis is that the hydrogen model produced goes to a strong simplification, where the fine and hyperfine structure of the hydrogen spectrum are just the consequence of the interaction between the magnetic moments of the proton and the electron, dictated by the distance and angle between them.\\
The descriptions made so far \cite{yepez, yepez2} are in complete agreement with the arguments and hopes driven by superstring theorists. In this paper a Superstring theory consistent model of the hydrogen atom is offered.

\section{Postulates}
1) \textbf{The electron is not moving}\\
After being trapped by the proton and the radiation of the energy quanta equivalent to the ionization potential has been emitted, \textbf{the electron is not moving any longer}. When another quantum transition, involving the absorption or emission of a photon happens, the electron just increases or decreases in size producing the next energy state, just as current quantum theories explain with the invention of the orbit.\\
2) \textbf{The shape of the electron is the intersection of a hypertorus in spaceland}\\
The different shapes of the orbits (the electron) are the consequence of the intersection of a hypertorus (4D-toroid) with a three dimensional space (See appendix A). The magnetic moment vector departs from the geometric center of the hypertorus.  These different intersections produce different magnetic momenta, which interaction with the proton's magnetic moment determines the fine and hyperfine structure of the hydrogen spectrum.\\
3) \textbf{The atom energy}\\
This is the sum of the following energies:\\
a) \textbf{The mass energy}: the masses of the proton and the electron repel each other and therefore this repulsion goes against the coulomb attraction. This energy has been wrongly designated as the kinetic energy of a point particle in current hydrogen models.\\
b) \textbf{The coulomb energy}: which produces the attraction between the proton and the electron and would produce the fusion of both particles (the collapse) unless the mass energy prevent it.\\
c) \textbf{The magnetic energy}: which results from the coupling of the magnetic moments of the proton and the electron and it is the sole responsible for the fine and hyperfine structure of the hydrogen spectrum.\\

\section{The Model}
Two mobi\"ous string currents embedded in a spiral-toroidal dimension just as it has been described in \cite{yepez} intersects the plane, leaving electric field vectors with different orientations, resulting in a wave pattern when the object is moving (this is described by the Dirac equation). These electric field vectors use all the dimensions of the space they are intersecting, producing a tension on such space and curving it: this phenomenon is the particle's mass (see Appendix B). The intersection event consumes some time, therefore the charge of the electron occurs. Also, these electric field vectors happen in a toroidal intersecting path, producing an anapole moment and a magnetic moment depending in how the space is intersected. In the case of the hydrogen atom, depending on the way that the electron's hypertorus intersects the space, different magnetic momenta are produced.

\subsection{Mass Energy}
Main energy levels for the hydrogen atom has been derived in the Bohr model, assuming the quantization of the orbital angular momentum \cite{eisberg130},
\begin{equation}
L=mvr= n\hbar
\end{equation}
and the mechanical stability of the orbit, 
\begin{equation}
\frac{1}{4\pi\epsilon_{0}}\frac{e^{2}}{r^{2}}=m\frac{v^{2}}{r}
\end{equation}
solving for \textit{v} in (1) and substituting in (2) yield the Bohr radius,
\begin{equation}
r=4\pi\epsilon_{0}\frac{n^{2}\hbar^{2}}{me^{2}}
\end{equation}
However, the same result can be obtained without the quantization of the angular momentum. By invoking the uncertainty principle, where $\triangle x=R$, the atom radius, 
\begin{equation}
\triangle p\triangle x=\hbar=>\triangle p=\frac{\hbar}{R}
\end{equation}
and  making $\triangle$p=p, the kinetic energy is, therefore, 
\begin{equation}
K=\frac{p^{2}}{2\mu}=\frac{(\triangle p)^{2}}{2\mu}=\frac{\hbar^{2}}{2\mu R^{2}}
\end{equation}
where $\mu$ is the reduced mass of the atom. The energy of the atom is, finally \cite{eisberg295},
\begin{equation}
E = K + V =\frac{\hbar^{2}}{2\mu R^{2}}-\frac{e^{2}}{4\pi\epsilon_{0}R}
\end{equation}
The Bohr radius is obtained again when the minimum energy in equation (6) is obtained through derivation,
\begin{equation}
\frac{dE}{dR}=-\frac{\hbar^{2}}{\mu R^{3}}+\frac{e^{2}}{4\pi\epsilon_{0}R^{2}}=0=>R=r=4\pi\epsilon_{0}\frac{ \hbar^{2}}{me^{2}}
\end{equation}
Equation (6) provides the main energy levels of the hydrogen atom. From postulate 1, equation (6) should be a potential energy instead of kinetic. Equation (4) becomes,
\begin{equation}
m \frac{\triangle x}{\triangle t}\triangle x= m \frac{\triangle x^{2}}{\triangle t}=\frac{\hbar}{2}
\end{equation}
From the uncertainty between energy and time,
\begin{equation}
\triangle E \triangle t = \frac{\hbar}{2}=> \triangle t = \frac{\hbar}{2 \triangle E}
\end{equation}
Substituting (9) in (8) gives,
\begin{equation}
m \frac{\triangle x^{2}}{\frac{\hbar}{2 \triangle E}}=\frac{\hbar}{2} => \triangle E = \frac{\hbar^{2}}{4 m \triangle x^{2}}
\end{equation}
making $\triangle E=E$ and $\triangle x= \frac{1}{\sqrt{2}}r$ the first part of equation (6) is obtained. If the product of uncertainties is $\hbar$, then $\triangle x = \sqrt{2} r$ to have the first part of (6). The same result was obtained with geometrical arguments departing from the ratio between $r_{\theta}$ and $r_{\varphi}$ of the deuteron \cite{yepez2}. Thus, the elimination of time as a variable frees the path for any other kind of energy, since postulate 1 forbids kinetic energy, this energy is designated as the mass energy of the particle. To obtain the different energy levels of the hydrogen atom, the square of the principal quantum number should multiply the first part of (6). Since there is no particle movement the reduced mass should not be used and the first part of equation (6) has to be separated for each particle,
\begin{equation}
E = M_{e}+ M_{p} + V =\frac{n^{2}\hbar^{2}}{2m_{e} r^{2}}+\frac{n^{2}\hbar^{2}}{2m_{p} r^{2}}-\frac{e^{2}}{4\pi\epsilon_{0}r}
\end{equation}
where M is the mass energy of the electron and the proton respectively. Equation (11) provides the main energy levels of the hydrogen atom.\\

\subsection{Magnetic Energy}
According to postulate 2 the electron magnetic moment would interact with the proton's magnetic moment at the very empty center of the atom, regardless of the way in which both particles are intersecting the 3-D space. Hence, the magnetic energy will come from the coupling of the magnetic moments of the proton and the electron (see Fig. 1). This energy is attractive and of the form \cite{magnetomechanics},
\begin{figure}
\begin{center}
\includegraphics[width=1in,height=2.5in]{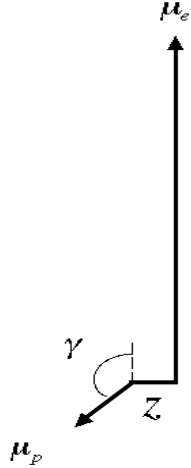}
\caption{Coupling (no to scale) between the magnetic moment of the electron and the magnetic moment of the proton}
\end{center}
\end{figure}
\begin{equation}
E_{z} = \frac{\mu_{0}\mu_{e}\mu_{p}(cos\gamma + sin\gamma)}{4\pi z^{3}}
\end{equation}
where $\mu_{0}$ is the permeability of the vacuum, $\mu_{e}$ and $\mu_{p}$ are the magnetic momentum of the electron and the proton respectively, $\gamma$ is the angle between both vectors and \textit{z} is the separation between the origin of the vectors.\\
When the magnetic moment vectors involved in the hydrogen atom meet at its geometric center  (with a separation \textit{z}$\sim 0.01 fm$), they do so at such an angle as to maintain an energy minimum, such a minimum happens only at two angle values: 135 and 315$^{\circ}$. At these angles the fourth term in equation (14) is in the order of $10^{-9} eV$. However, a \textit{z} distance in the order quoted above will produce the energies of the hyperfine structure. A direct consequence of this mechanism is that there are just two states of magnetic coupling between the proton and the electron, see Fig. 1 and 4.\\
Therefore, the equation that describe the total energy of the hydrogen atom is, 
\begin{equation}
E =\frac{n^{2}\hbar^{2}}{2m_{e} r^{2}}+\frac{n^{2}\hbar^{2}}{2m_{p} r^{2}}-\frac{e^{2}}{4\pi\epsilon_{0}r}-\frac{\mu_{0}\mu_{e}\mu_{p}(cos\gamma + sin\gamma)}{4\pi z^{3}}
\end{equation}
by making $r_{\theta}=\frac{\hbar}{2 m c}$ and the atomic radius being $r_{\varphi}$ of the torus (see ref \cite{yepez2}), equation (13) becomes,
\begin{equation}
E =\frac{n^{2}r_{\theta, e}\hbar c}{r_{\varphi}^{2}} + \frac{ n^{2}r_{\theta, p}\hbar c}{r_{\varphi}^{2}} -\frac{e^{2}}{4\pi\epsilon_{0}r_{\varphi}}-\frac{\mu_{0}\mu_{e}\mu_{p}(cos\gamma + sin\gamma)}{4\pi z^{3}}
\end{equation}
According to postulate 3, the natural repulsion between the mass energy fields of the proton and the electron does not allow the later to have its total mass energy ($m_{e}c^{2}$ = 0.5 MeV, which happens at $r_{\varphi}$ = 273 \textit{fm}) because the electric attraction is not enough to compensate for the repulsion between the masses of both particles. On the contrary, such repulsion makes the electron expands, reducing its mass energy to 13.6 eV at 53000 \textit{fm}. Given that the attractive coulomb energy at that radius is -27.2 eV, a minimum of -13.6 eV is produced, see Fig. 2.\\
This reduction in the mass energy of the electron, however, only happens because the attachment to the proton and it is a state in its mass energy field described by its confinement. The free particle mass $\sim$ 500000 eV is recuperated when the electron is free and can shrink back again to a $r_{\varphi}$ = 273 \textit{fm} because the proton is not there anymore. Therefore, the total mass energy of the electron is conserved as a field in either case. As it can be noticed, the electron mass is no longer a fixed scalar, but a field described by its confinement. 

\begin{figure}
\begin{center}
\includegraphics[width=5in,height=7in]{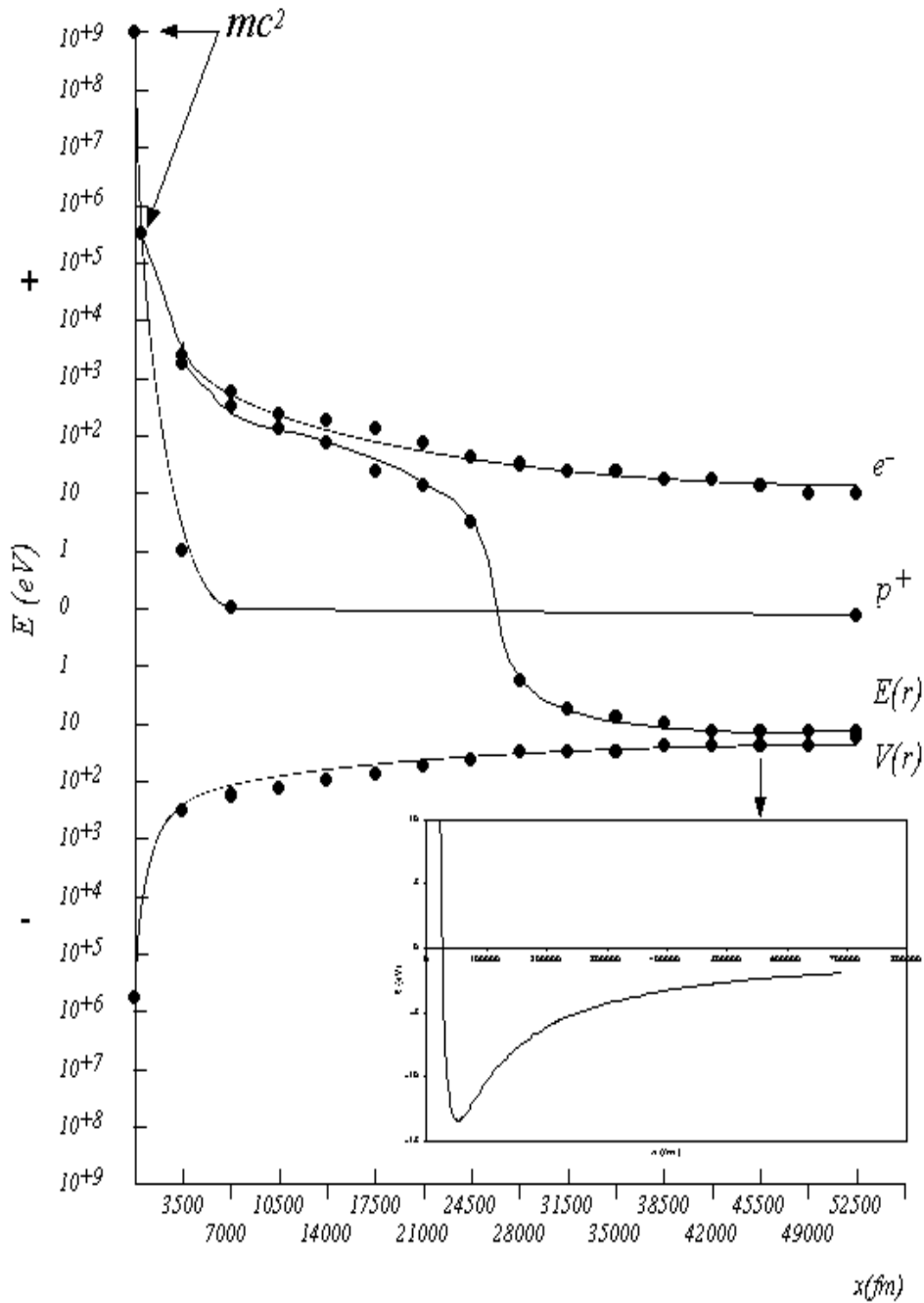}
\caption{Mass and coulomb energies (equation 11) vs atomic radius for the hydrogen atom. The minimum occurs at the Bohr radius (insert), The mass energy field of the proton still contribute with 7 meV at that radius.}
\end{center}
\end{figure}

\subsection{Magnetic Moment and Charge of the Free electron}
As it was discussed for the flat deuteron \cite{yepez2}, the charge of the flat electron occurs at the intersection time $\frac{\theta r_{\theta}}{c}$, i.e.,
\begin{equation}
e= i_{\theta}\frac{\theta r_{\theta}}{c}=> i_{\theta}=\frac{ec}{\theta r_{\theta}}
\end{equation}
The fraction of spire that intersect the plane will be the arc $\theta r_{\theta}$ in the whole $\theta$ perimeter $2\pi r_{\theta}$,
\begin{equation}
\eta= \frac{\theta}{2\pi}
\end{equation}
therefore, the flat electron's anapole moment will be,
\begin{equation}
T= \frac{1}{4\pi c}\cdot IV= \frac{1}{4\pi c} \frac{\theta}{2\pi} \frac{ec}{\theta r_{\theta}} 2\pi^2r_{\theta}^2r_{\varphi}= \frac{e r_{\theta}r_{\varphi}}{4}
\end{equation}
Since the intersection time is $\frac{\theta r_{\theta}}{c}$, the magnetic moment will be,
\begin {equation}
\mu=\frac{\frac{e r_{\theta}r_{\varphi}}{4}}{\frac{\theta r_{\theta}}{c}} = \frac{ec r_{\varphi}}{4\theta}
\end{equation}
for the free particle, $r_{\varphi}=\sqrt{2}r_{\theta, e}$, $\theta=\frac{\sqrt{2}}{4}$ to obtain the electron's magnetic moment,
\begin {equation}
\mu= \frac{ec\sqrt{2}r_{\theta, e}}{4\frac{\sqrt{2}}{4}}=\frac{ec\hbar}{2 m_{e}c}=\frac{e \hbar}{2 m_{e}}
\end{equation}

\subsection{Magnetic Moment of the Trapped electron}
It was found that when the electron is trapped by the proton to form a hydrogen atom, the intersection time changes to  $\frac{\theta r_{\theta}}{\alpha c}$, where $\alpha$ is $\frac{e^2}{2h\epsilon_{0}}$,
\begin{equation}
e= i_{\theta}\frac{\theta r_{\theta}}{\alpha c}=> i_{\theta}=\frac{e \alpha c}{\theta r_{\theta}}
\end{equation}
therefore, the flat electron's anapole moment will be,
\begin{equation}
T= \frac{1}{4\pi c}\cdot IV= \frac{1}{4\pi c} \frac{\theta}{2\pi} \frac{e \alpha c}{\theta r_{\theta}} 2\pi^2r_{\theta}^2r_{\varphi}= \frac{e \alpha r_{\theta}r_{\varphi}}{4}
\end{equation}
Since the intersection time is $\frac{\theta r_{\theta}}{\alpha c}$, the magnetic moment will be,
\begin {equation}
\mu=\frac{\frac{e \alpha r_{\theta}r_{\varphi}}{4}}{\frac{\theta r_{\theta}}{\alpha c}} = \frac{e \alpha^2 c r_{\varphi}}{4\theta}
\end{equation}

\subsection{Shapes and the transitions in spaceland}
As it was fully described in \cite{yepez2}, the shapes of the deuteron can be understood as the different ways of intersecting a hypertorus with a 3D space (see appendix A). Those shapes are: a sphere inside another sphere (\textit{ss}), a torus (\textit{t}) and two separated spheres (\textit{ts}). These experimental shapes highly resemble an \textit{S}, a $P_{3/2}$ and a $P_{1/2}$ orbitals respectively (compare \cite{eisberg300} with \cite{forest}). Hence, it is postulated that the electron orbitals are the different ways of intersection of a hypertorus with a 3D space, i.e. postulate 2.\\
The hyperfine structure of the hydrogen atom, require that the shape \textit{ss} (an \textit{S} orbital) has a magnetic moment and according to ref \cite{yepez2}, such shape should not have it. This apparent contradiction happens because the model of intersection between a 3D body and a 2D plane (flatland) is still far from the real model which consists of a 4D body intersecting a 3D space. Therefore, it has to be proposed that  in the real case, the state \textit{ss} (an S like state) will have both kind of momenta. In other words, the anapole moment only happens when such toroidal dipole moment aims into time (according to the flatland model) and therefore is not consuming it, otherwise it will be a magnetic moment. Therefore, the transitions between different shapes of the electron will go through a no magnetic moment (anapole moment) state, see Fig. 3.
\begin{figure}
\begin{center}
\includegraphics[width=4in,height=2in]{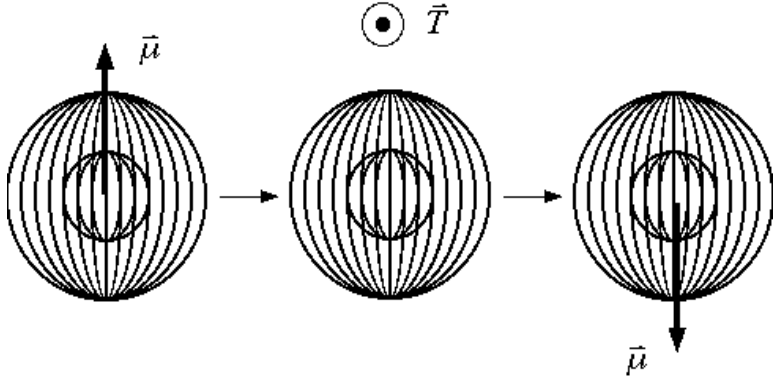}
\caption{Transition in between energy levels for an S like intersection of the electron in the hydrogen atom.}
\end{center}
\end{figure}

\section{Results}
As it can be clearly appreciated in ref \cite{Phydrogen}, each electron orbital (energy state or electron way's of intersection with the space) presented different slopes in their respective plots of energy (frequency) versus magnetic field strength for the hyperfine structure of the hydrogen spectrum. This means that the electron magnetic moment is changing between the different kind of intersections. For a \textit{ss} intersection (\textit{S} orbital) the magnetic moment measured was $\mu_{e}$, but for the \textit{ts} intersection ($P_{1/2}$) the magnetic moment of the electron diminished to $\frac{\mu_{e}}{\pi}$. Surprisingly, the \textit{t} intersection state ($P_{3/2}$) showed two different magnetic moments, one at $\frac{2\mu_{e}}{\pi}$ and the other that doubles the normal magnetic moment of the free electron, i.e. 2 $\mu_{e}$. Table 1 presents the angle $\theta$ needed in equation (22) to reproduce the experimental magnetic moment observed for each energy state.
\begin{figure}
\begin{center}
\includegraphics[width=3.5in,height=2.5in]{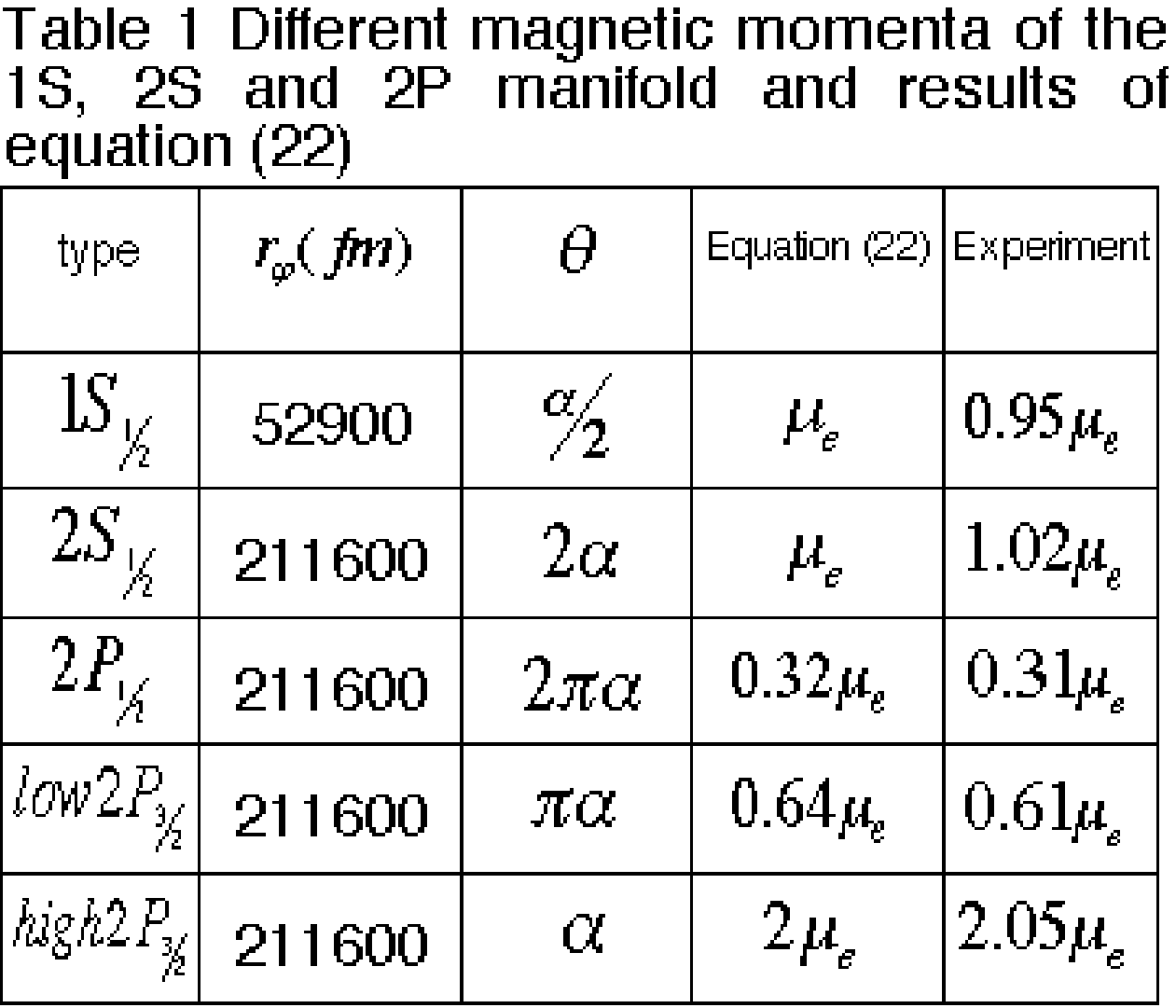}
\end{center}
\end{figure}Also, it is observed in \cite{Phydrogen} that for the \textit{ss} (S orbital) and \textit{ts} intersections ($P_{1/2}$), there are four signals: three of them depart from the higher energy part of the hyperfine transition and one departs from the lower energy part.\\
\begin{figure}
\begin{center}
\includegraphics[width=6in,height=7in]{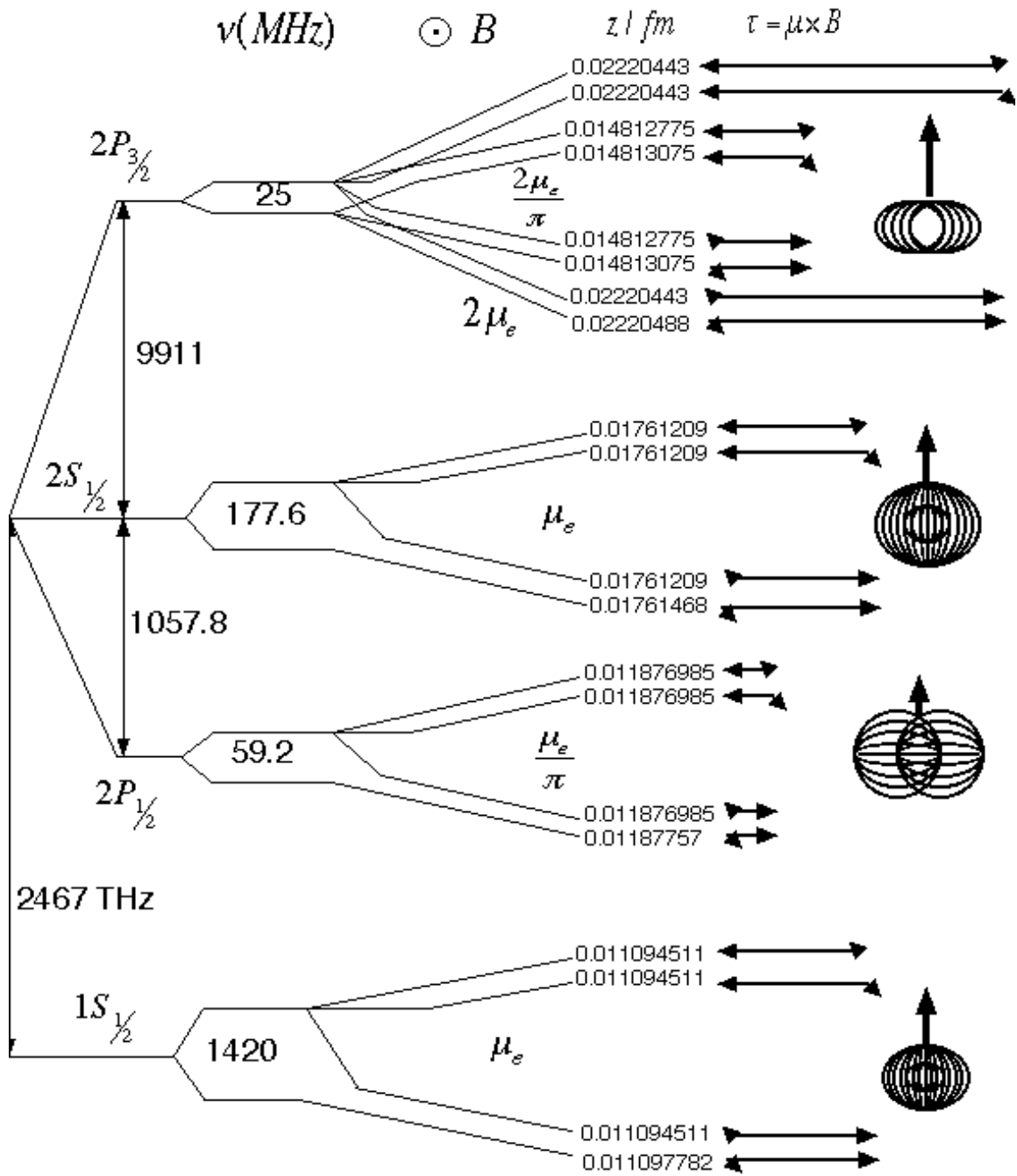}
\caption{Hydrogen Energy levels for states with principal quantum numbers n=1 and n=2.}
\end{center}
\end{figure}
In Fig. 4, the total energy levels of the hydrogen atom with principal quantum number n=1 and n=2 (according to the model presented in this paper) is shown. The Lamb shift, transition $2S_{1/2}-2P_{1/2}$ is 1057845.0 (9.0) kHz \cite{Lundeen}, the $2S_{1/2}-2P_{3/2}$ is 9911200 kHz \cite{Hagley}, the 1S hyperfine frequency (hff) is 1420.4057517667 MHz \cite{Richard}, the 2S hff is 177556800 Hz \cite{Karshenboim}. With the later value the $2P_{1/2}$ hff was calculated to be 59185595 Hz following the scheme in ref \cite{Heberle}. The $2P_{3/2}$ hff was measured from  \cite{Phydrogen} to be 25 MHz. Then, the theoretical magnetic momenta (Table 1) and the \textit{z} distance needed to give the corresponding energy state  are presented. Fig. 4 can be explained as follow: The 4D torus electron intersects the 3D space in different ways and with different magnetic moments. Each magnetic moment has two orientations with respect to the proton's magnetic moment, which also has two orientations possible. As a result, the combination of both particles magnetic moments and orientations produces four energy states for each magnetic moment produced. The number of ways the electron can intersect the space depends on the principal quantum number, when n=1 just an \textit{ss} (S orbital) intersection occurs: three of the four possible energy states have a relative shorter \textit{z} distance making them relatively more energetics and therefore, they occur in the upper part of the hyperfine energy state. However, one of this group of three necessarily has an electron magnetic moment orientation that will make it a lower energy seeker. Hence, even though this energy state is a relative higher energy state at zero magnetic field strength, it will go toward lower energies as the magnetic field strength increases as it is found experimentally \cite{Phydrogen}. When n=2 the \textit{ss} intersection increases in size and maintains the same magnetic moment, $\mu_{e}$. Such increase in size changes the main energy level of such state. But the same figures as in the lower quantum number state happens. The higher quantum number, however, opens the possibility of other ways of intersection with other magnetic moments. For instance, the intersection \textit{ts} presents the lowest magnetic moment for this quantum number and therefore the lowest energy state in the family of intersections. Surprisingly enough, the intersection \textit{t} occurs with a magnetic moment that double the magnetic moment of the free electron and with another that is two times the magnetic moment of the intersection \textit{ts}.

\section{Discussion}
When the electron is trapped by the electric field of the proton, it expands because the mass energy of the proton repels it, thus it reduces its mass energy to a minimum. Then, the electric attraction acts to compress the electron, until the natural repulsion between the mass energy fields of the two particles prevents further compression. The electron does compress in steps, so the movement of its charge from one step to the other produces a vibration, which translates to an energy releasing process: the emission of an energy quantum ($h\nu$). In this act, a photon or sum of photons equivalent to the ionization potential occurs. This process is highly reversible and the absorption of photons makes the electron expand against the electric attraction, decreasing its mass energy toward a minimum again, until the electron is free from the coulomb attraction and decreases its size, regaining the free particle size ($r_{\varphi}$=273 fm and $r_{\theta}$=193 fm).\\
As a result of the atom formation, the electron presents a reduced mass energy at the Bohr radius and due to the coulomb attraction, a minimum of energy occurs. This process is responsible for the stability of the atom. The electron can not present its total mass energy at the Bohr radius because the presence of the proton prevents the electron from reducing its size further. Its total mass energy is, therefore, warped space between both particles (see Fig. 2).\\
This relationship between mass and size, described by the mass energy equation (14) is consistent with the general theory of relativity, which states that the mass-energy tells the space how to curve: mass exerts a tension on the space, therefore, the smaller the mass the lesser the tension and the larger the space the particle will occupy. The mechanism behind such a tension on the space is the use of all the available dimensions by the electric field vectors upon their intersection and this depends on the structure of the string currents in a fermion or a photon. If such electric field vectors use  one dimension less than the available in such a space, no tension and thus no mass occurs, i.e. the photon \cite{yepez}. As it has been shown, this relationship is also derived from the uncertainty principle, which is fundamental in the quantum realm. Therefore, this relationship is the conceptual connection between both theories.\\
Nonetheless, there is experimental evidence that supports "a larger than a proton electron" \cite{compton, compton1}. The classic idea that the electron should be smaller than the  proton because it has a smaller mass, have prevailed precisely because the "success" of point like particle models (Bohr model and subsequences). This success, however, has diminished progressively as the accuracy of the measurements in the hydrogen spectrum increased and those models are now uncapable to explain the hyperfine structure of the hydrogen spectrum.\\
Postulate 1 requires not using the kinetic energy as the cause of the atom stability,  postulate 2 is needed to explain the different shapes of the electron (orbits in the current understanding), which leads to different magnetic momenta and the soul of the interaction that finally explains the hyperfine interaction: the coupling between those magnetic momenta of the electron and the proton (see Fig.4). Therefore, the third postulate is a consequence of the first two.\\
The shapes of the electron are related to the different ways of intersecting space \cite{yepez2} and, as it was shown in Table 1, with different magnetic momenta. This later finding being an experimental fact \cite{Phydrogen}. Probably, the proton, like the electron, also has different magnetic momenta and shapes in its domain but for the sake of simplicity the "normal" magnetic moment of the proton was kept constant for the calculations.\\
The author wonders about the strong similarities between the shape of the orbitals derived from the Schr\"odinger solution of the hydrogen atom and the intersection of a hypertorus with a 3D space (the later is equivalent to the revolution of the intersection of a 3D-torus with a plane), because the former is the consequence of the solutions of the colatitude equation \cite{eisberg298} and the latter comes from geometric arguments \cite{yepez2}. As a matter of fact, this resemblance was used to identify each toroidal intersection with a given energy state. However, Schr\"odinger's treatment does not say anything about the magnetic moment of the different orbitals and/or establish that an orbital does have a magnetic moment and tell less about its direction. Moreover, since postulate 1 prevents a wave as the solution for the hydrogen atom, the same shape structures can be obtained with the revolution of Cassini ovals, which are the mathematical expression of the intersection of a torus by a plane \cite{cassini}. Surprisingly this have been done. Cassini ovals has been used to model two center electron orbitals \cite{bolotin}. Therefore, the model presented is self-consistent. Given that it also describes the behavior of the wave when the electron is moving \cite{yepez}, the present model is a particle-wave model for real.\\
Another strong similarity is the statistical interpretation of the present point particle electron, which can be anywhere in the orbit as a smeared out distribution of mass and charge. The model presented here produces its charge in the act of intersection with the space, thus the charge of the electron will appear and disappear from existence all over its intersection shape; and in a way that the charge will be \textit{e} at any measurable time allowed by the uncertainty principle. On the other hand, the mass is the tension exerted by the electric field vectors, which also happens concurrently with the charge. Thus, in all these aspects, both models coincide.\\
The model presented has also a strong relation with string theory since two mobi\"ous string currents in a spiral-toroidal extra-dimension are used to built the same. Therefore, it would be very interesting to see which of the string theories is consistent with this model.\\
Fig. 4 shows, clearly, that both the fine and hyperfine structure of the spectrum is just the consequence of the coupling of different magnetic moments at different distances with the proton and those distances are at the very empty center of the proton. This explanation departs completely from the current understandings and unify fine and hyperfine structure as being different aspects of the same process, contrary to current understanding which attributes the fine structure to the spin-orbit interaction and the hyperfine to the coupling between the proton and the electron's spin.\\
The transition in the majority of the states consist just in the inversion of the electron magnetic moment, keeping the separation between it and the proton's magnetic moment, constant. There is only one state that changes its \textit{z} distance; this explains why there are three signals in the upper part of each hyperfine frequency and only one in the lower part. Thus, the whole hydrogen spectrum can be explained with the primary "big energy" electron size and mass transition and the secondary "little energy" electron-proton magnetic moment coupling.\\
According to the Shr\"odinger solution of the hydrogen atom, a \textit{S} orbital should have less energy than a \textit{P} orbital. However, as it can be observed in Fig. 4, the appearance of different magnetic moments depending on the way the electron is intersecting the space changes this previous understanding, where the magnetic moment of the electron did not play any role.
Thus, the fact that the $2S_{1/2}$ energy state has more energy than the $2P_{1/2}$ (the Lamb shift) is no longer a mystery since the electron's magnetic moment of the former is higher than the later.\\
Finally, the author is in complete agreement with the idea that it is impossible to imagine the electron, because the only thing one can see is its multiple shadows.
 
\section{Conclusions}
A simpler model of the atom has been achieved. This model is free of non-radiating moving particles and thermodynamics contradictions. It uses the natural coupling of the magnetic moment of the particles involved in the atom to explain the hyperfine structure of the hydrogen spectrum. The electron \textbf{is} the orbit and its magnetic moment departs from its geometrical center. This moment changes according to the way it intersects the space and therefore its coupling with the proton magnetic moment presents different energy states concordantly. The working hypothesis: "the electron mass is the tension that it exerts on the space, thus its mass energy decreases when the electron occupies a larger space" is perfectly consistent with the General Theory of Relativity and with Quantum Mechanics.

\section{Acknowledgment}

The author fully appreciate the efforts of John Byrne and Constantino Badra for useful grammar corrections of the text.

\section{Appendix A, Kind of Intersection of a hypertorus with a 3 D space, the magnetic and the anapole moment of the electron does not commute}

\begin{figure}
\begin{center}
\includegraphics[width=3in,height=3in]{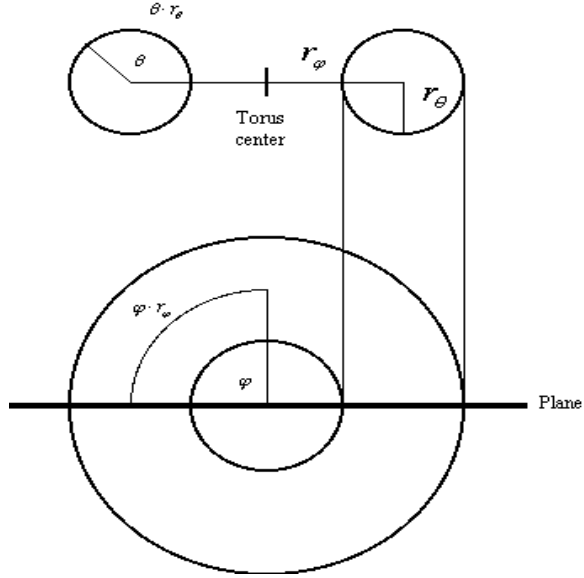}
\caption{Cross section of a 3D Torus intersecting a plane (bottom) and its orthogonal projection (above),  showing its geometric parameters r$_{\theta}$ and r$_{\varphi}$. The arcs $\varphi r_{\varphi}$ and $\theta r_{\theta}$ are also shown. }
\end{center}
\end{figure}

The technique to look at the intersections of a 4D object in a 3D space consist in making the intersection of its 3D version with a 2D space (a plane) and then revolving the resultant through a given edge \cite{beyond the third}. For example, if a 4D sphere moves through a 3D space, its intersection will begin as a tiny sphere that increases in size up to a certain limit (the radius of the 4D sphere) and then reduces its size until it stops intersecting such space. The equivalent operation between a 3D sphere and a 2D plane will be a tiny circle that increases up to a maximum size (the radius of the sphere) and then reduces its size until it disappears from the plane. Therefore, it is easy to visualize that the revolution of this circle will give the equivalent sphere of the previous "impossible to imagine" operation. This is true for the sphere which is highly symmetrical, as any given rotation edge will give the same result. In the case of a 4D torus intersecting the 3D space, however, there are a wide number of possible intersections depending on the angle between the torus and the plane. However, one is only concerned with the possible intersections that gives the femtometer toroidal structures found for the deuteron in reference \cite{forest} and these are obtained with just two orientations: $\varphi$ and $\theta$ orientations.\\
In Fig. 5 the cross section of the intersection of a torus with a plane and its orthogonal projection are depicted, the upper half of the torus is in the future and the bottom half in the past. In this figure the torus is in its $\theta$ orientation, which means that the plane is cutting the torus along this angle and in a symmetric way. The $\varphi$ orientation is obtained when the torus intersects the plane through that angle in a symmetric way also. Fig. 6 presents the result of the operation described in the previous paragraph, for a torus that is intersecting the plane in its $\theta$ and $\varphi$ orientations.
Given that the string currents travel through the shape of the toroidal dimension, a torodial charge moment called anapole moment is produced. Electric field vectors are left in the plane following such structure during the act of intersection, a intersection time is consumed and a charge is produced regardless to the orientations of the intersection. However, there is a fundamental difference between these two situations: when the anapole moment is perpendicular to the plane ($\varphi$ orientation) and when it is in the plane ($\theta$ orientation). In the $\varphi$ orientation the anapole moment is aiming into time and thus, it is not consuming it. In the $\theta$ orientation such moment is in the plane, consumes time and becomes a magnetic moment. Therefore, anapole and magnetic momenta should not commute.
\begin{figure}
\begin{center}
\includegraphics[width=4in,height=5in]{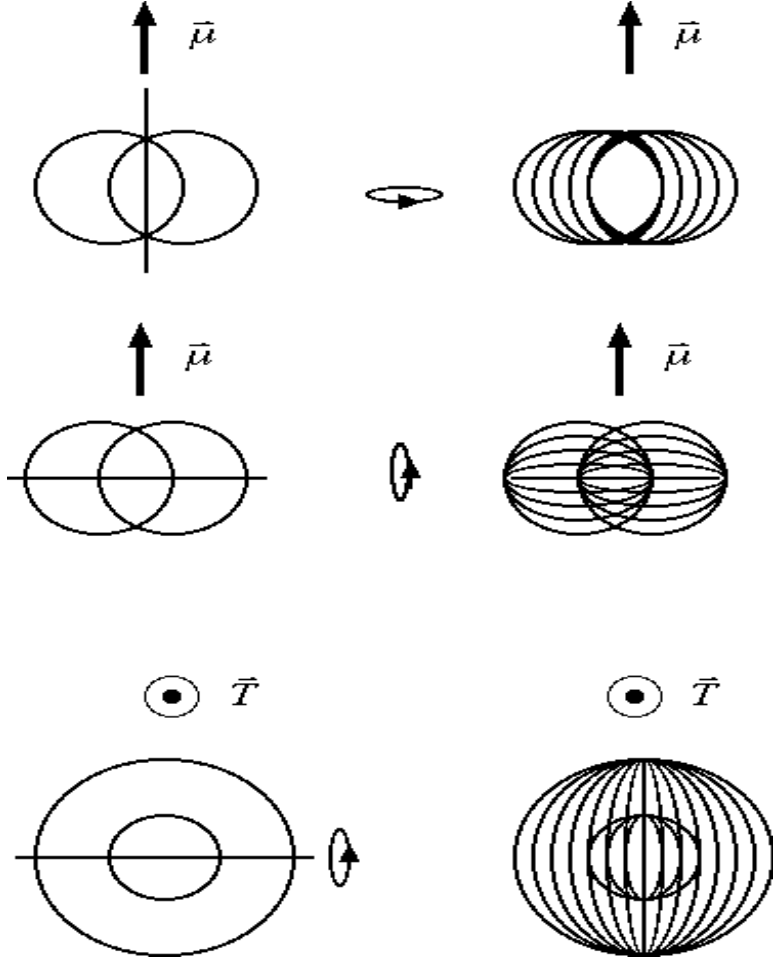}
\caption{Intersection of a 3D torus with the plane (the paper sheet) and the resultant revolution through a given edge. The first two figures comes from a $\theta$ orientation and shape at the bottom from a $\varphi$ orientation. The revolution of these projections is on the right side of the figure. Such revolution are exactly the shapes found for the deuteron in ref. \cite{forest}.}
\end{center}
\end{figure}

\section{Appendix B, The Concept of Mass}

On Fig. 7 and 8 the string like structure of the electron and photon are clearly observed. The pink planes represent our observable universe, the spiral-toroidal dimension is depicted in grey and the string current is shown as a series of red dots with blue arrows. Previously \cite{yepez}, it was thought that the current travelled all along the spiral dimension; however, this makes impossible the charge conservation in EPR paradox (Appendix C), because the $\varphi$ intersection does not leave the charge \textit{e} at the fixed time $\frac{\theta r_{\theta}}{c}$ as the $\theta$ orientation does. According to Fig. 7, it is proposed that Inertia is due to the counter movement detected for the electron in its $\varphi$ axis, because the electron has counter movements between its both toroids and  the photon does not show this characteristic, thus photon does not has inertia as it is required. Also, the counter movement detected is the intrinsic way to make the matter wave produced to be dependent upon the speed of the particle. In the case of the photon, both toroids move in the same $\varphi$ direction and therefore, its wave has a fixed speed, i.e. inertia is a requirement for the matter wave to happen.

\begin{figure}
\begin{center}
\includegraphics[width=3.3in,height=7.2in]{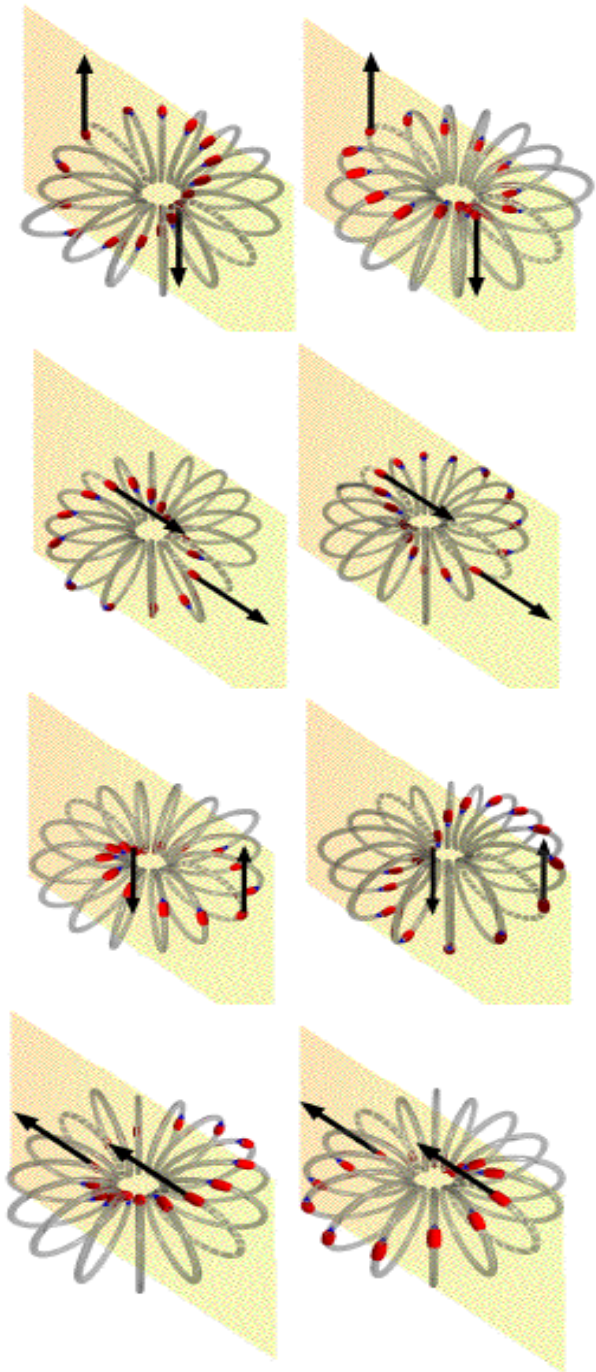}
\caption{Flat electron intersection with 2D space (flatland) sequence, propagation is downward the page.}
\end{center}
\end{figure}

\begin{figure}
\begin{center}
\includegraphics[width=3.3in,height=7.2in]{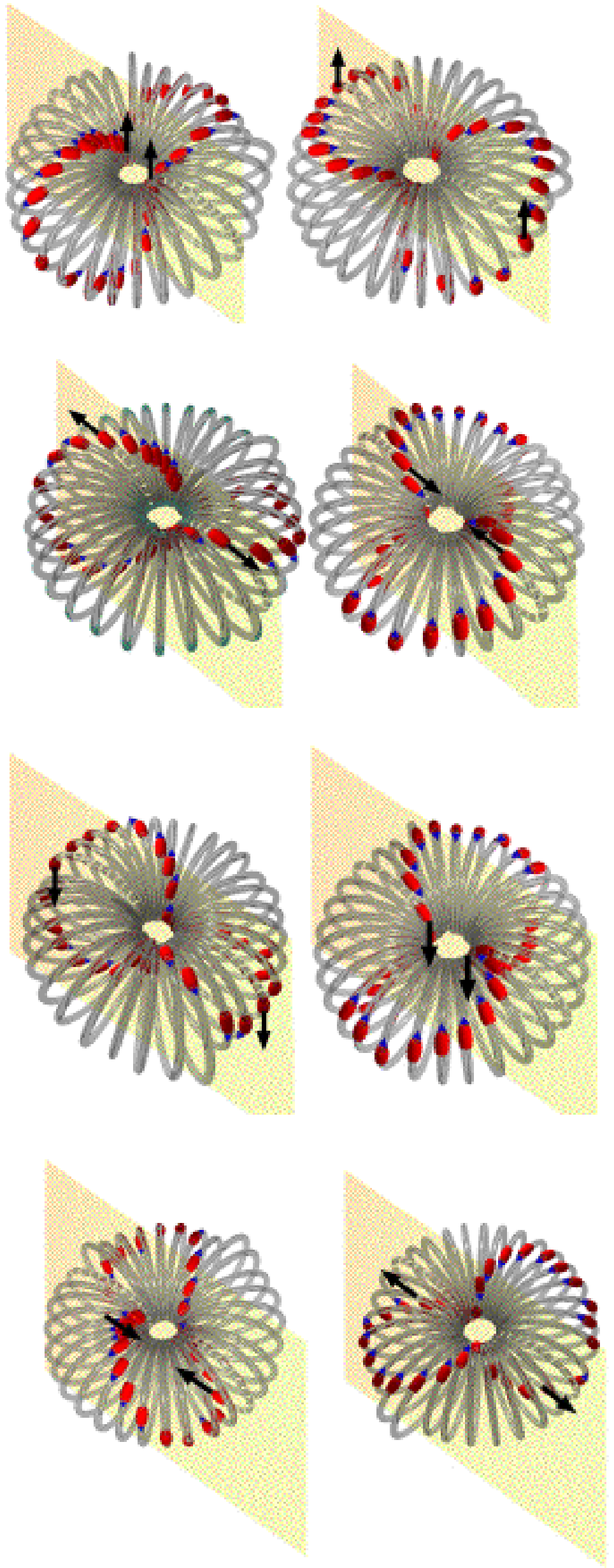}
\caption{Flat photon intersection with 2D space (flatland) sequence, propagation is downward the page.}
\end{center}
\end{figure}

\begin{figure}
\begin{center}
\includegraphics[width=6in,height=7in]{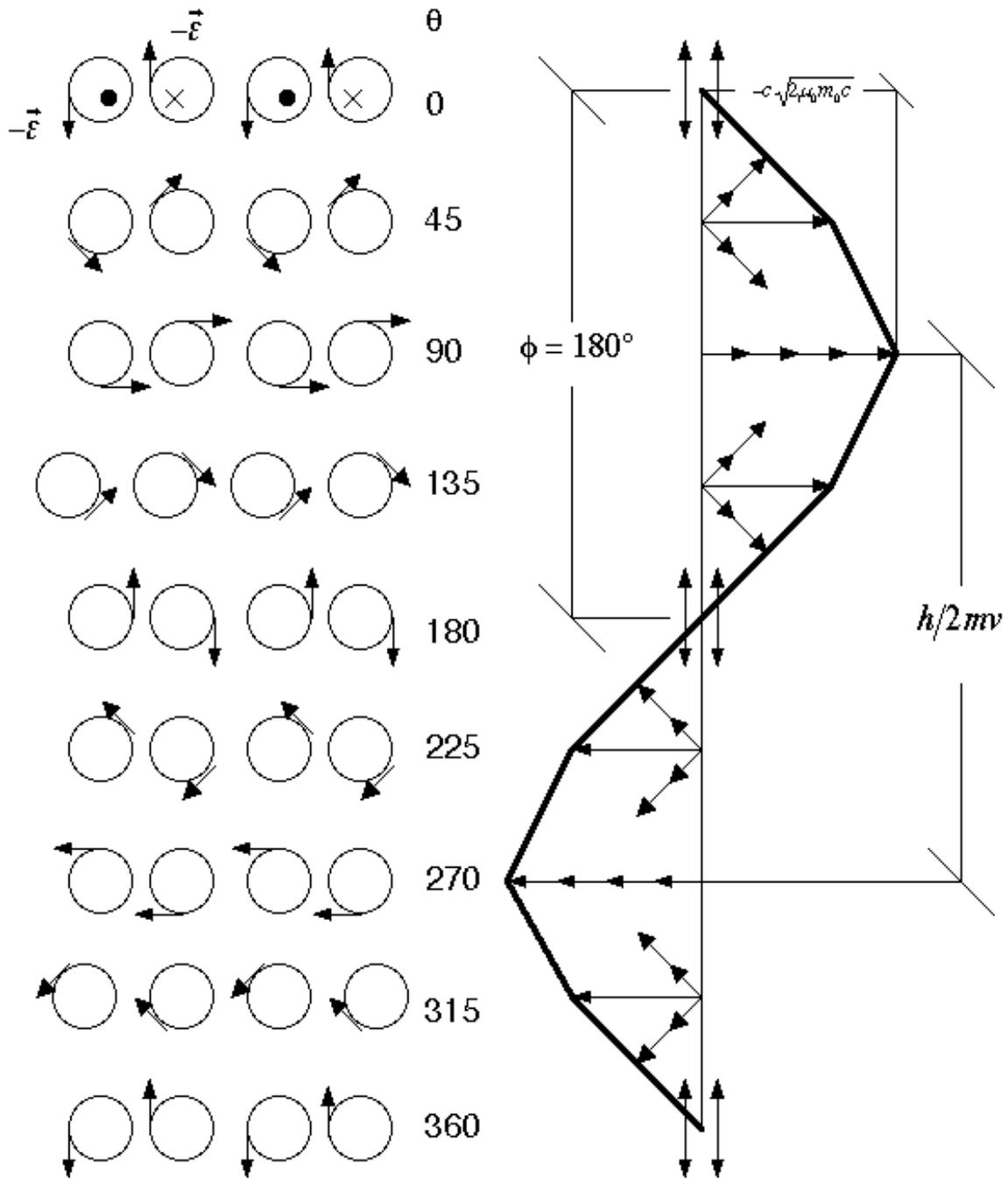}
\caption{Sinusoidal Electric Field Printed in the Plane by the flat electron. Maximum and minimum go perpendicular to the direction of movement, which is downward the page.}
\end{center}
\end{figure}

\begin{figure}
\begin{center}
\includegraphics[width=5.5in,height=7in]{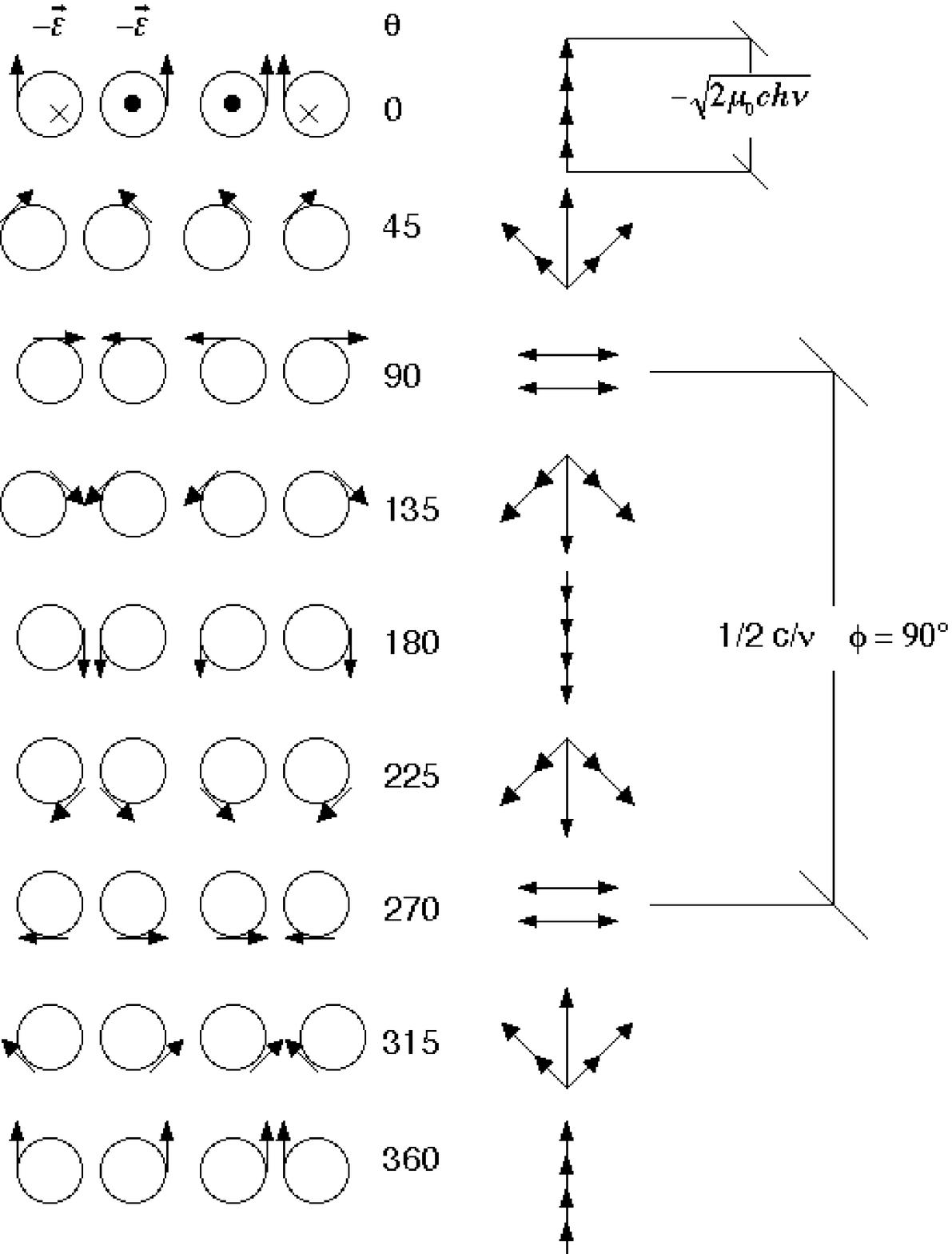}
\caption{Sinusoidal Electric Field Printed in the Plane by flat photon. Maximum and minimum go in the direction of movement, which is downward the page.}
\end{center}
\end{figure}

Fig. 9 and 10 present the track left by the electron and the photon. As the electron moves through flatland, its electric field vectors occur perpendicularly to the direction of movement, thus the wave generated oscillates using all the dimensions of the paper sheet: one dimension to move and the other to oscillate. In the case of the photon, its electric field vectors go with the direction of the movement, thus the wave generated uses one dimension less than the two available. Given that the photon requires not to have mass, it was concluded that in order to have mass the object should use all the dimensions of the space it is intersecting. As the electron's electric field vectors do just that, a tension is exerted into the space and this is the particle mass.\\
On Fig. 11 the force produced by an accelerated electron is appreciated. As acceleration progressed, the electron's DeBroglie wavelength diminished and therefore more intersections events will happen per unit of space. As a result, a progressively higher bending of the space occurs, i.e. a force \footnote{force is the effect produced by the geometry of the space-time, if there is no bending of the space there is momentum but no force is produced}. And this is why force is the product of the mass per acceleration of the object. As the photon does not tenses the space, a change in its wavelength is irrelevant with respect to the shape of the space-time and it just follows whatever that shape is. This is why the photon just has momentum as $P=\frac{E}{c^{2}}$ and does not produce a force. If the electron is not accelerated, the same amount of intersection events per unit of space would happen, the space is not progressively tensed and in this case the electron just has momentum, as the product of its mass per its velocity. Hence, the electron can have momentum and force but the photon can  only have momentum. A consequence of this concept is that the particle mass can no longer be treated as a fixed scalar but as a dependable field. 

\begin{figure}
\begin{center}
\includegraphics[width=5.5in,height=5in]{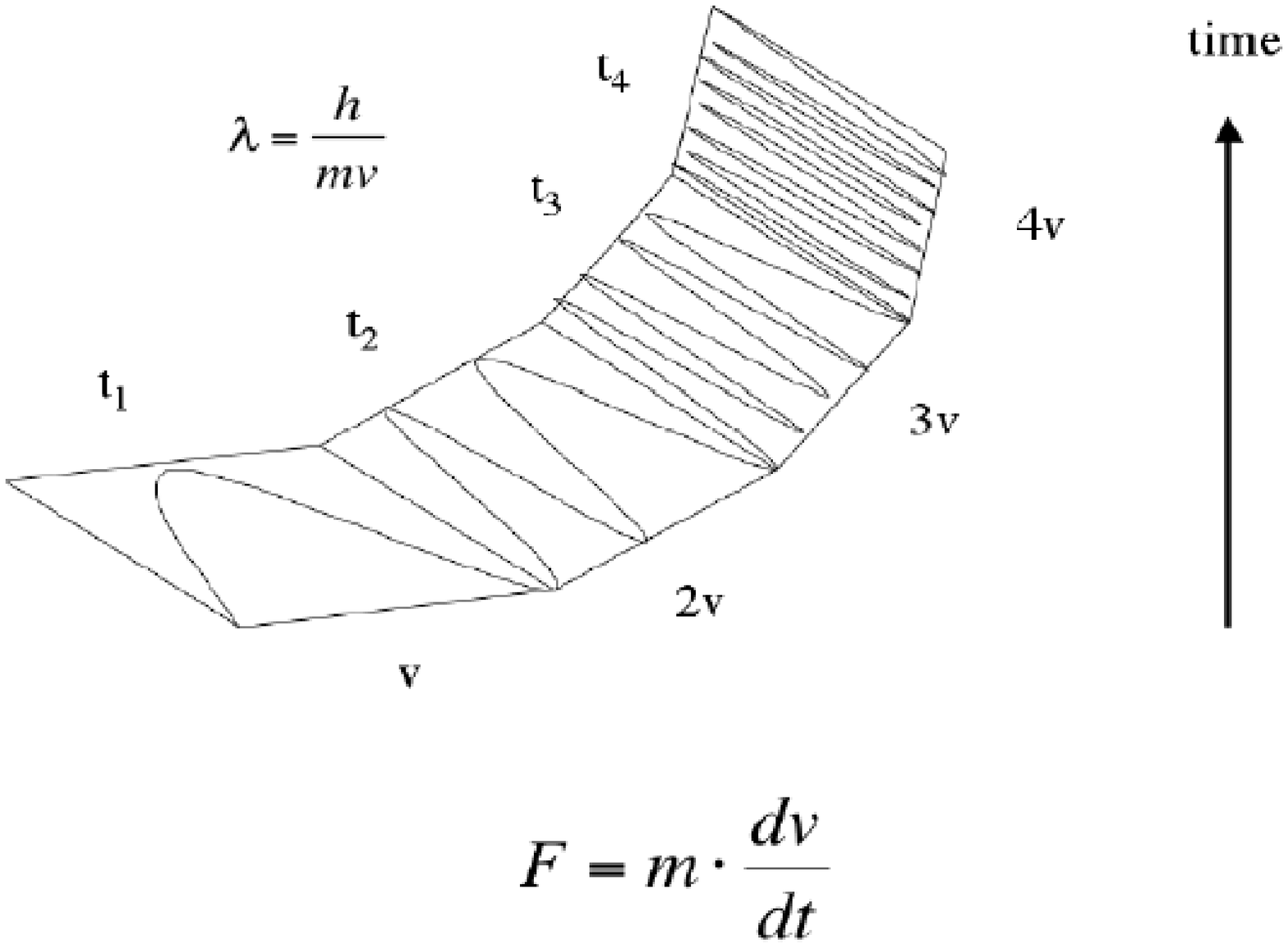}
\caption{Force because an accelerated matter wave}
\end{center}
\end{figure}

\section{Appendix C, EPR paradox}

On Fig. 12, the flatland equivalence of the Stern-Gerlach experiment for flat electrons is observed. Since that universe has only two dimensions, a moving particle can go just  to the right or left, or can be accelerated or retarded. In this case, a very clear separation between up and down oriented magnetic dipole moments to the left and to the right is observed.\\
Fig. 13 is an example of the travel of flat electrons through different oriented magnetic filters in flatland. After passing the first right-filter, the change in orientation of the second filter to 90$^{\circ}$, made the electron to intersect flatland at its $\varphi$ axis. This process is at random, therefore, the orientation produced by the first filter is lost. However, the number of electrons is reduced to a half, because in this orientation half of the electrons are accelerated and the other half are retarded (the blocking device is out of the plane). Also, this orientation produces a lost in the identification of the particle because its magnetic dipole moment is out of the plane (aiming into time). Finally, upon another random rearrangement of the 8 accelerated electrons in a left-filter, 4 electrons with their magnetic dipole moment completely opposed to the direction of the magnetic dipole moment of the 16 electrons which passes the first filter, occurs. And this behavior is consistent with the 3D experimental fact. \\
The change in the orientation of the intersection from $\theta$ to $\varphi$ like, accompanied with the 50:50 stochastic process is revealing in regard with the uncertainty principle. The $\theta$ orientation is what produces the sinusoidal electric field pattern printed in the plane and it is the way to know the momentum of the particle with precision. However, to do so the particle is in two places of the plane at the same time, therefore, the measurement of its exact position is even nonsensical. It is pure non-local to produce the wave that leaves in the plane. The stochastic change to the $\varphi$ intersection stops the particle being non-local, to occupy just one place in flatland, thus its position can be measured very precisely. But in doing so, it stops also to produce the sinusoidal pattern, ergo its momentum is completely unknown and there is no way to avoid this, because it is a particle-space intrinsic property. Clearly, this is why position and momentum do not commute.
\begin{figure}
\begin{center}
\includegraphics[width=5.5in,height=7in]{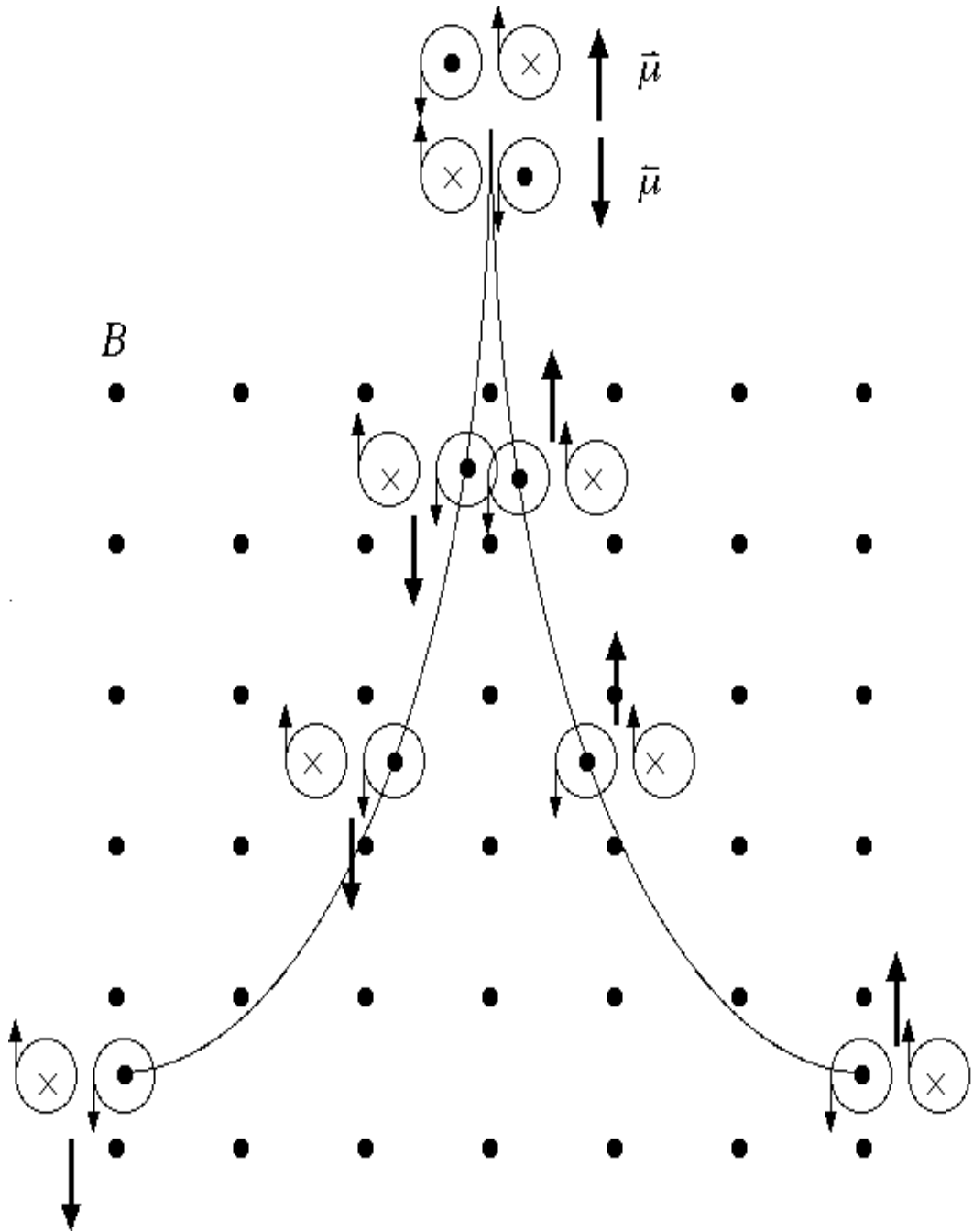}
\caption{Stern-Gerlach experiment in flatland.}
\end{center}
\end{figure}
\begin{figure}
\begin{center}
\includegraphics[width=5.5in,height=8in]{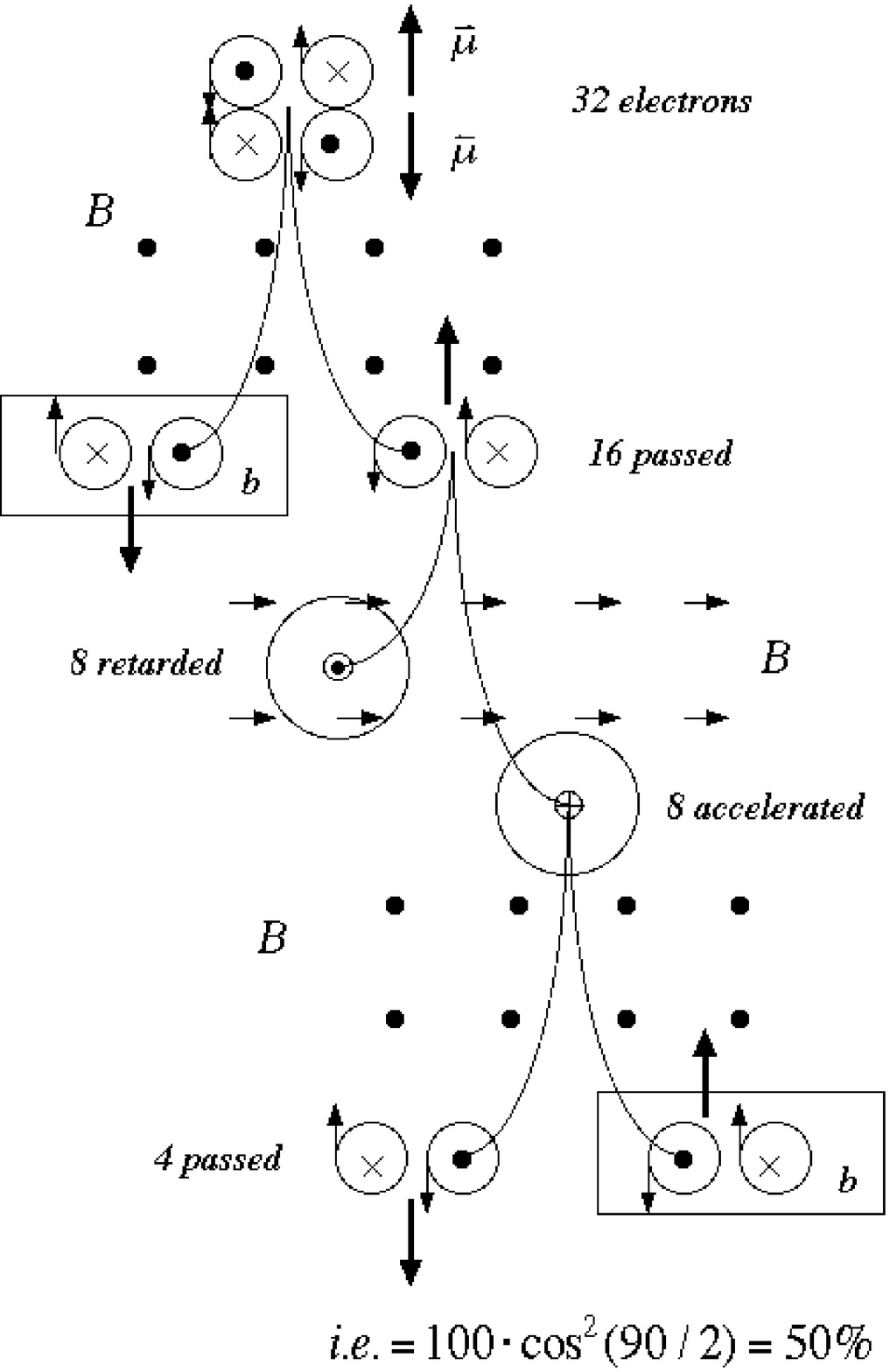}
\caption{Flat electrons through magnetic filters with different orientations}
\end{center}
\end{figure}


\begin{thebibliography}{1}

\bibitem{eisberg277}
R. Eisberg and R. Resnick,
 \textit{ÒF\'{\i}sica Cu\'antica, \'atomos, mol\'eculas, s\'olidos, n\'ucleos y part\'{\i}culasÓ} 
(Mexico: Limusa, 1999), p. 277.

\bibitem{chemist electron}
T. Arabatzis and K. Gavroglu,
 \textit{Eur. J. Phys.}, 
18 (1997) 150.

\bibitem{Hspectrum1}
Oxford Library of the Physical Sciences, G.W. Series,
 \textit{ÒSpectrum of Atomic HydrogenÓ,} 
(Oxford University Press, 1957), p. 41.

\bibitem{eisberg297}
R. Eisberg and R. Resnick,
 \textit{ÒF\'{\i}sica Cu\'antica, \'atomos, mol\'eculas, s\'olidos, n\'ucleos y part\'{\i}culasÓ} 
(Mexico: Limusa, 1999), p. 297.

\bibitem{hyperspace}
Michio Kaku,
\textit{ Hyperspace, A Scientific Odyssey Through Parallel Universes, Time Warps, and the 10$^{th}$ Dimensions}
Anchor Books, New York, p. 342.

\bibitem{yepez}
O. Y\'epez, "Matter and Light in Flatland"
\textit{arXiv:physics/}
0401153 v1 28 Jan 2004.

\bibitem{eisberg130}
R. Eisberg and R. Resnick,
 \textit{ÒF\'{\i}sica Cu\'antica, \'atomos, mol\'eculas, s\'olidos, n\'ucleos y part\'{\i}culasÓ} 
(Mexico: Limusa, 1999), p. 130.

\bibitem{eisberg295}

 \textit{Ibid,} 
295.

\bibitem{magnetomechanics}
Hull J. R.,
 \textit{Supercond. Sci. Technol.}
13 (2000) 854.

\bibitem{yepez2}
O. Y\'epez, "The Flatland Deuteron"
\textit{arXiv:physics/}
0510168 v1 18 Oct 2005.

\bibitem{eisberg300}
R. Eisberg and R. Resnick,
 \textit{ÒF\'{\i}sica Cu\'antica, \'atomos, mol\'eculas, s\'olidos, n\'ucleos y part\'{\i}culasÓ} 
(Mexico: Limusa, 1999), p. 300.

\bibitem{forest}
J.L. Forest, V.R. Pandharipande, S.C. Pieper, R.B. Wiringa, R. Schiavilla and A. Arriaga, 
  \textit{Phys. Rev. C}, 
 54 (1996) 646.

\bibitem{Phydrogen}
G.F. Bassani, M. Inguscio and T.W. H\"ansch,
 \textit{ÒProceedings, The Hydrogen AtomÓ} 
(Springer-Verlag Berlin Heidelberg 1989), p. 120.

\bibitem{Lundeen}
S.R. Lundeen and F.M. Pipkin,
\textit{Metrologia}
22 (1986) 9.

\bibitem{Hagley}
E. W. Hagley and F.M. Pipkin,
\textit{Phys. Rev. Lett.}
72 (1994) 1172.

\bibitem{Richard}
R. Ley, D.  Hagena, D. Weil, G.Werth,
\textit{Hyperfine Interaction 89 (1994) 327}
187 (1990) 145.
\bibitem{Karshenboim}
S.G. Karshenboim, P. Fendel, V. G. Ivanov, N. N. Kolachevsky, T. W. H\"ansch,
  \textit{Can. J. Phys.}, 
 83 (2005) 283.

\bibitem{Heberle}
J. W. Heberle, H. A. Reich and P. Kusch,
  \textit{Phys. Rev.}, 
 101 (1956) 612.

\bibitem{compton}
A.H. Compton,
  \textit{Phys. Rev.}, 
14 (1919) 20.

\bibitem{compton1}

  \textit{Ibid}, 
14 (1919) 247.

\bibitem{cassini}
E. W., Weisstein,
  \textit{"Cassini Ovals." From MathWorld--A Wolfram Web Resource}, 
 http://mathworld.wolfram.com/CassiniOvals.html

\bibitem{bolotin}
A.B. Bolotin, A.V. Kazimianec,
 \textit{Lithuanian Journal of Physics} 
35 (1995) 346.

\bibitem{eisberg298}
R. Eisberg and R. Resnick,
 \textit{ÒF\'{\i}sica Cu\'antica, \'atomos, mol\'eculas, s\'olidos, n\'ucleos y part\'{\i}culasÓ} 
(Mexico: Limusa, 1999), p. 298.

\bibitem{beyond the third}
Thomas F. Banchoff,
 \textit{"Beyond the Third Dimension, Geometry, Computer Graphics, and Higher Dimensions"}
(New York, Scientific American Library, 1996), Chapter three.

\end{thebibliography}
 \end{document}